\documentclass{article}

\usepackage{amsmath}
\usepackage{amsthm}
\usepackage{amssymb}
\usepackage{graphicx}
\usepackage{verbatim}
\usepackage{natbib}
\usepackage{caption}
\usepackage{subcaption}
\usepackage{fancyvrb}
\usepackage{enumerate}
\usepackage{relsize}
\usepackage{algorithm}
\usepackage{algpseudocode}

\usepackage[svgnames]{xcolor}
\usepackage{listings}

\usepackage{hyperref}
\usepackage[margin=1.2in]{geometry}
\hypersetup{colorlinks,citecolor=blue,urlcolor=blue,linkcolor=blue}

\usepackage{stefan_tex}
\graphicspath{{./figures/}}

\begin{document}

\title{Estimating Treatment Effects with Causal Forests: \\An Application}
\author{ Susan Athey \\ \texttt{athey@stanford.edu}
       \and
        Stefan Wager \\ \texttt{ swager@stanford.edu}}
\date{Stanford University}

\maketitle

\begin{abstract}
We apply causal forests to a dataset derived from the National Study of Learning Mindsets, and consider resulting practical and conceptual challenges. In particular, we discuss how causal forests use estimated propensity scores to be more robust to confounding, and how they handle data with clustered errors.
\end{abstract}


\section{Methodology and Motivation}

There has been considerable recent interest in methods for heterogeneous treatment effect estimation
in observational studies \citep*{athey2015machine,athey2018generalized,ding2016randomization,
dorie2017automated,hahn2017bayesian,hill2011bayesian,imai2013estimating,kunzel2017meta,luedtke2016super,
nie2017learning,shalit2017estimating,su2009subgroup,wager2017estimation,zhao2017selective}.
In order to help elucidate the drivers of successful approaches to treatment effect estimation,
Carlos Carvalho, Jennifer Hill, Avi Feller and Jared Murray organized a workshop at the 2018
Atlantic Causal Inference Conference and asked several authors to analyze a shared dataset
derived from the National Study of Learning Mindsets \citep{yeager2016using}.

This note presents an analysis using causal forests \citep*{athey2018generalized,wager2017estimation};
other approaches will be discussed in a forthcoming issue of \emph{Observational Studies}
with title ``Empirical Investigation of Methods for Heterogeneity.'' All analyses are carried out using
the \texttt{R} package \texttt{grf}, version \texttt{0.10.2} \citep{grf,R}.
Full replication files are available at \texttt{github.com/grf-labs/grf},
in the directory \texttt{experiments/acic18}.

\subsection{The National Study of Learning Mindsets}
\label{sec:data}

The National Study of Learning Mindsets is a randomized study conducted in U.S.
public high schools, the purpose of which was to evaluate the impact of a
nudge-like intervention designed to instill students with a growth
mindset\footnote{According to the National Study, ``A growth mindset is the belief that
intelligence can be developed. Students with a growth mindset understand they can get smarter
through hard work, the use of effective strategies, and help from others when needed.
It is contrasted with a fixed mindset: the belief that intelligence is a fixed trait that is set in stone
at birth.''} on student achievement. To protect student privacy, the present analysis is not based
on data from the original study, but rather on data simulated from a model fit to the
National Study dataset by the workshop organizers. The present
analysis could serve as a pre-analysis plan to be applied to the original
National Study dataset \citep{nosek2015promoting}.

\begin{table}
\begin{center}
\begin{tabular}{cp{0.8\textwidth}}
\hline
\texttt{S3} & Student's self-reported expectations for success in the future, a proxy for prior achievement, measured prior to random assignment \\
\texttt{C1} & Categorical variable for student race/ethnicity \\
\texttt{C2} & Categorical variable for student identified gender \\
\texttt{C3} & Categorical variable for student first-generation status, i.e. first in family to go to college \\
\texttt{XC} & School-level categorical variable for urbanicity of the school, i.e. rural, suburban, etc. \\
\texttt{X1} & School-level mean of students' fixed mindsets, reported prior to random assignment \\
\texttt{X2} & School achievement level, as measured by test scores and college preparation for the previous 4 cohorts of students \\
\texttt{X3} & School racial/ethnic minority composition, i.e., percentage of student body that is Black, Latino, or Native American \\
\texttt{X4} & School poverty concentration, i.e., percentage of students who are from families whose incomes fall below the federal poverty line \\
\texttt{X5} & School size, i.e., total number of students in all four grade levels in the school \\ \hline
$Y$ & Post-treatment outcome, a continuous measure of achievement \\
$W$ &Treatment, i.e., receipt of the intervention \\ \hline
\end{tabular}
\caption{Definition of variables measured in the National Study of Learning Mindsets}
\label{tab:variable_def}
\end{center}
\end{table}

Our analysis is based on data from $n = 10,391$ children from a probability sample of $J = 76$
schools.\footnote{Initially, 139 schools were recruited into the study using
a stratified probability sampling method \citep{gopalan2018national}. Of these
139 recruited schools, 76 agreed to participate in the study; then, students were
individually randomized within the participating schools. In this note, we
do not discuss potential bias from the non-randomized selection of 76 schools among
the 139 recruited ones.}
For each child $i = 1, \, ..., \, n$, we observe a binary treatment indicator $W_i$,
a real-valued outcome $Y_i$, as well as 10 categorical or real-valued covariates
described in Table \ref{tab:variable_def}. We expanded out categorical random variables
via one-hot encoding, thus resulting in covariates $X_i \in \RR^p$ with $p = 28$.
Given this data, the workshop organizers expressed particular interest in the three
following questions:
\begin{enumerate}
\item Was the mindset intervention effective in improving student achievement?
\item Was the effect of the intervention moderated by school level achievement (\texttt{X2}) or pre-existing mindset norms (\texttt{X1})? In particular there are two competing hypotheses about how \texttt{X2} moderates the effect of the intervention: Either it is largest in middle-achieving schools (a ``Goldilocks effect'') or is decreasing in school-level achievement.
\item Do other covariates moderate treatment effects?
\end{enumerate}
We define causal effects via the potential outcomes model \citep{imbens2015causal}: For each
sample $i$, we posit potential outcomes $Y_i(0)$ and $Y_i(1)$ corresponding to the outcome
we would have observed had we assigned control or treatment to the $i$-th sample, and
assume that we observe $Y_i = Y_i(W_i)$. The average treatment effect is then defined as
$\tau = \EE{Y_i(1) - Y_i(0)}$, and the conditional average treatment effect function is
$\tau(x) = \EE{Y_i(1) - Y_i(0) \cond X_i = x}$.

\begin{figure}
\begin{centering}
\includegraphics[width=0.45\textwidth]{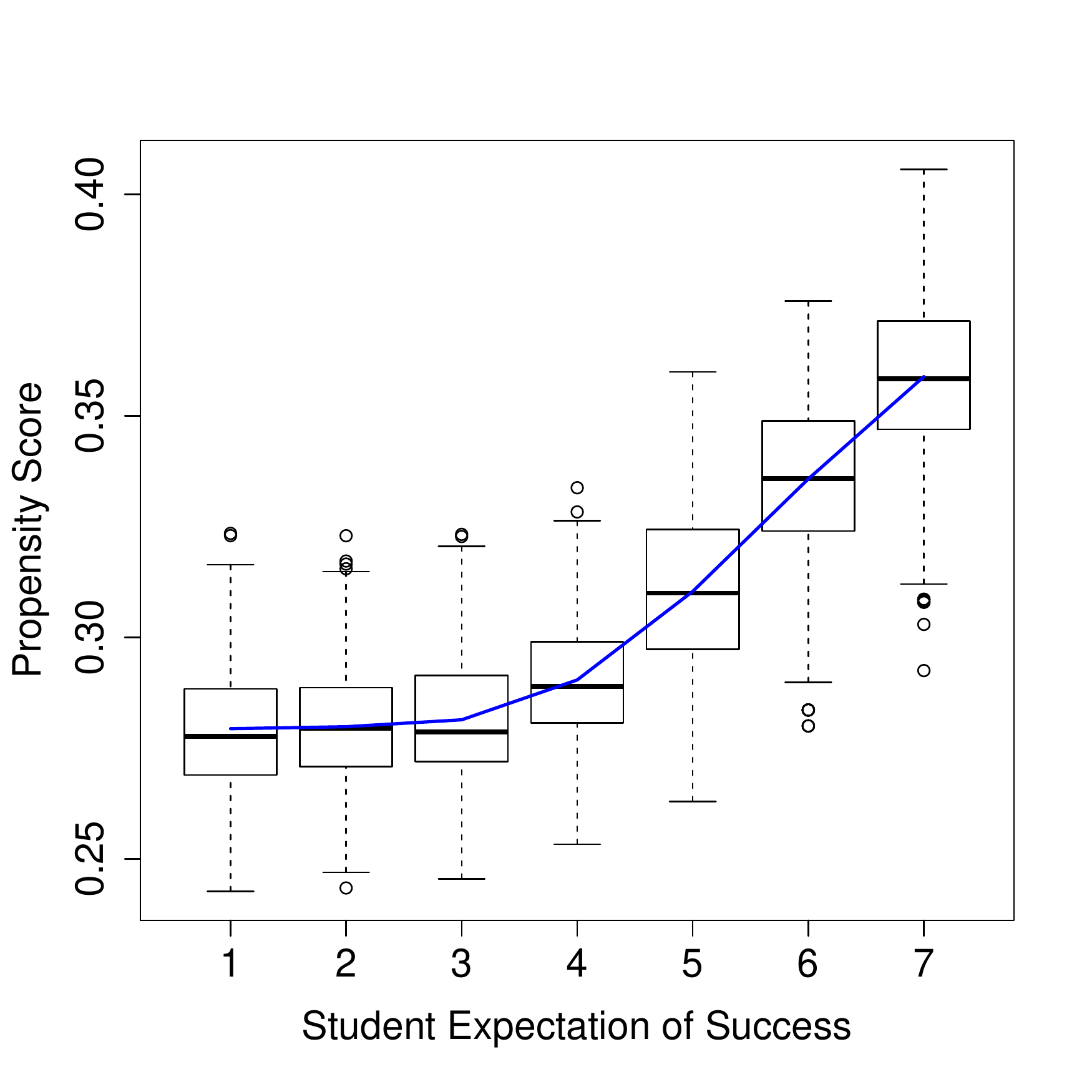}
\caption{Visualizing estimated treatment propensities against student expectation of success.}
\label{fig:pre}
\end{centering}
\end{figure}

This dataset exhibits two methodological challenges. First,
although the National Study itself was a randomized study, there seems to be some selection
effects in the synthetic data used here. As seen in Figure \ref{fig:pre},
students with a higher expectation of success appear to be more likely to receive treatment.
For this reason, we analyze the study as an observational rather than randomized study. In order
to identify causal effects, we assume unconfoundedness, i.e., that treatment assignment is as
good as random conditionally on covariates \citep{rosenbaum1983central}
\begin{equation}
\label{eq:unconf}
\cb{Y_i(0), \, Y_i(1)} \indep W_i \cond X_i.
\end{equation}
To relax this assumption, one could try to find an instrument for treatment
assignment \citep{angrist2008mostly}, or conduct a sensitivity analysis for hidden 
confounding \citep{rosenbaum2002observational}.

Second, the students in this study are not independently sampled; rather, they are all drawn from
76 randomly selected schools, and there appears to be considerable heterogeneity across schools.
Such a situation could arise if there are unobserved school-level features that are important treatment
effect modifiers; for example, some schools may have leadership teams who implemented the intervention
better than others, or may have a student culture that is more receptive to the treatment.
If we want our conclusions to generalize outside of the 76 schools we ran the experiment in,
we must run an analysis that robustly accounts for the sampling variability of potentially unexplained school-level effects.
Here, we take a conservative approach,
and assume that the outcomes $Y_i$ of students within a same school may be arbitrarily
correlated within a school (or ``cluster''), and then apply cluster-robust analysis tools
\citep*{abadie2017should}.

The rest of this section presents a brief overview of causal forests, with an emphasis of how they
address issues related to clustered observations and selection bias.
Causal forests are an adaptation of the random forest algorithm of \citet{breiman2001random}
to the problem of heterogeneous treatment effect estimation. For simplicity, we start below by
discussing how to make random forests cluster-robust in the classical case of non-parametric
regression, where we observe pairs $(X_i, \, Y_i)$ and want to estimate
$\mu(x) = \EE{Y_i \cond X_i = x}$. Then, in the next section, we review how forests
can be used for treatment effect estimation in observational studies.

\subsection{Cluster-Robust Random Forests}
\label{sec:clust}

When observations are grouped in unevenly sized clusters, it is important to carefully define
the underlying target of inference. For example, in our setting, do we want to fit a model that
accurately reflects heterogeneity in our available sample of $J = 76$ schools, or one that we hope will
generalize to students from other schools also? Should we give more weight in our analysis
to schools from which we observe more students?

Here, we assume that we want results that generalize beyond our $J$ schools, and that we give each
school equal weight; quantitatively, we want models that are accurate for predicting effects on a new
student from a new school. Thus, if we only observed outcomes $Y_i$ for students with school membership
$A_i \in \cb{1, \, ..., \, J}$ we would estimate the global mean as \smash{$\hmu$} with standard
error \smash{$\hsigma$}, with
\begin{equation}
\label{eq:cluster_mu}
\hmu_j = \frac{1}{n_j} \sum_{\cb{i : A_i = j}} Y_i, \ \ \ \
\hmu = \frac{1}{J} \sum_{j = 1}^J \hmu_j, \ \ \ \ 
\hsigma^2 = \frac{1}{J(J-1)} \sum_{j = 1}^J \p{\hmu_j - \hmu}^2,
\end{equation}
where $n_j$ denotes the number of students in school $j$. Our challenge is then to use random
forests to bring covariates into an analysis of type \eqref{eq:cluster_mu}. Formally, we seek to carry
out a type of non-parametric random effects modeling, where each school is assumed to have some
effect on the student's outcome, but we do not make assumptions about its distribution (in particular,
we do not assume that school effects are Gaussian or additive).

At a high level, random forests make predictions as an average of $b$ trees, as follows:
(1) For each $b = 1, \, ..., \, B$, draw a subsample $\set_b \subseteq \cb{1, \, ..., \, n}$;
(2) Grow a tree via recursive partitioning on each such subsample of the data; and
(3) Make predictions
\begin{equation}
\label{eq:forest_pred}
\hmu(x) = \frac{1}{B} \sum_{b = 1}^B \sum_{i = 1}^n \frac{Y_i \, \1\p{\cb{X_i \in L_b(x), \, i \in \set_b}}}{\abs{\cb{i : X_i \in L_b(x), \, i \in \set_b}}},
\end{equation}
where $L_b(x)$ denotes the leaf of the $b$-th tree containing the training sample $x$.
In the case of out-of-bag prediction, we estimate \smash{$\hmu^{(-i)}(X_i)$} by only considering those
trees $b$ for which $i \not\in \set_b$. This short description of forests of course leaves many
details implicit. We refer to \citet{biau2016random} for a recent overview of random forests and
note that, throughout, all our forests are ``honest'' in the sense of \citet{wager2017estimation}.

When working with clustered data, we adapt the random forest algorithm as follows. In step (1), rather
than directly drawing a subsample of observations, we draw a subsample of clusters
$\jj_b \subseteq \cb{1, \, ..., \, J}$; then, we generate the set $\set_b$ by drawing $k$ samples
at random from each cluster $j \in \jj_b$.\footnote{If $k \leq n_j$ for all $j = 1, \, ..., \, J$, then each
cluster contributes the same number of observations to the forest as in \eqref{eq:cluster_mu}.
In \texttt{grf}, however, we also allow users to specify a value of $k$ larger than the smaller
$n_j$; and, in this case, for clusters with $n_j \leq k$, we simply use the whole cluster (without
duplicates) every time $j \in \jj_b$. This latter option may be helpful in cases where there are some
clusters with a very small number of observations, yet we want $\set_b$ to be reasonably large so
that the tree-growing algorithm is stable.} The other point where clustering matters is when we
want to make out-of-bag predictions in step (3). Here, to account for potential correlations within
each cluster, we only consider an observation $i$ to be out-of-bag if its cluster was not drawn in
step (1), i.e., if $A_i \not\in \jj_b$.

\subsection{Causal Forests for Observational Studies}
\label{sec:cf}

One promising avenue to heterogeneous treatment effect estimation starts from an early result of
\citet{robinson1988root} on inference in the partially linear model \citep*{nie2017learning,zhao2017selective}.
Write $e(x) = \PP{W_i \cond X_i = x}$ for the propensity
score and $m(x) = \EE{Y_i \cond X_i = x}$ for the expected outcome marginalizing over treatment.
If the conditional average treatment effect function is constant, i.e., $\tau(x) = \tau$ for all $x \in \xx$, then
the following estimator is semiparametrically efficient for $\tau$ under unconfoundedness \eqref{eq:unconf}
\citep{chernozhukov2017double,robinson1988root}:
\begin{equation}
\label{eq:robinson}
\htau = \frac{\frac{1}{n} \sum_{i = 1}^n \p{Y_i - \hatm^{(-i)}(X_i)}\p{W_i - \he^{(-i)}(X_i)}}{\frac{1}{n} \sum_{i = 1}^n \p{W_i - \he^{(-i)}(X_i)}^2},
\end{equation}
assuming that $\hatm$ and $\he$ are $o(n^{-1/4})$-consistent for $m$ and $e$ respectively in
root-mean-squared error, that the data is independent and identically distributed, and that we have
overlap, i.e., that propensities $e(x)$ are uniformly bounded away from 0 and 1. The $^{(-i)}$-superscripts
denote ``out-of-bag'' or ``out-of-fold'' predictions meaning that, e.g., $Y_i$ was not used to compute
\smash{$\hatm^{(-i)}(X_i)$}.

Although the original estimator \eqref{eq:robinson} was designed for constant treatment effect estimation,
\citet{nie2017learning} showed that we can use it to motivate an ``$R$-learner''
objective function for heterogeneous treatment
effect estimation,
\begin{equation}
\label{eq:rlearner}
\htau(\cdot) = \argmin_\tau \cb{\sum_{i = 1}^n \p{\p{Y_i - \hatm^{(-i)}(X_i)} - \tau(X_i)\p{W_i - \he^{(-i)}(X_i)}}^2 + \Lambda_n\p{\tau(\cdot)}},
\end{equation}
where $\Lambda_n\p{\tau(\cdot)}$ is a regularizer that controls the complexity of the learned $\htau(\cdot)$
function. A desirable property of this approach is that, if the true conditional average treatment
effect function $\tau(\cdot)$ is simpler than the main effect function $m(\cdot)$ or the propensity function
$e(\cdot)$, e.g., qualitatively, if $\tau(\cdot)$ allows for a sparser representation than $m(\cdot)$ or $e(\cdot)$,
then the function \smash{$\htau(\cdot)$} learned by optimizing \eqref{eq:rlearner} may converge faster than
the estimates for \smash{$\hatm(\cdot)$} or \smash{$\he(\cdot)$} used to form the objective function.

Causal forests as implemented in \texttt{grf} can be seen as a forest-based method motivated by the
$R$-learner \eqref{eq:rlearner}. Typically, random forests \citep{breiman2001random} are understood
as an ensemble method: A random forest prediction is an average of predictions made by individual
trees. However, as discussed in \citet*{athey2018generalized}, we can equivalently
think of random forests as an adaptive kernel method; for example, we can re-write the regression
forest from \eqref{eq:forest_pred} as
\begin{equation}
\label{eq:forest_kernel}
\hmu(x) = \sum_{i = 1}^n \alpha_i(x) Y_i, \ \ \alpha_i(x) =  \frac{1}{B} \sum_{b = 1}^B  \frac{\1\p{\cb{X_i \in L_b(x), \, i \in \set_b}}}{\abs{\cb{i : X_i \in L_b(x), \, i \in \set_b}}},
\end{equation}
where, qualitatively, $\alpha_i(x)$ is a data-adaptive kernel that measures how often the $i$-th training
example falls in the same leaf as the test point $x$.
This kernel-based perspective on forests suggests a natural way to use them for treatment
effect estimation based on \eqref{eq:robinson} and \eqref{eq:rlearner}:
First, we grow a forest to get weights $\alpha_i(x)$, and then set
\begin{equation}
\label{eq:cf}
\htau = \frac{\sum_{i = 1}^n \alpha_i(x) \p{Y_i - \hatm^{(-i)}(X_i)}\p{W_i - \he^{(-i)}(X_i)}}{\sum_{i = 1}^n \alpha_i(x) \p{W_i - \he^{(-i)}(X_i)}^2}.
\end{equation}
\citet*{athey2018generalized} discuss this approach in more detail, including how to design a
splitting rule for a forest that will be used to estimate predictions via \eqref{eq:cf}.
Finally, we address clustered observations by modifying the random forest sampling procedure
in an analogous way to the one used in Section \ref{sec:clust}.

Concretely, the \texttt{grf} implementation of causal forests starts by fitting two separate regression
forests to estimate \smash{$\hatm(\cdot)$} and \smash{$\he(\cdot)$}. It then makes out-of-bag
predictions using these two first-stage forests, and uses them to grow a causal forest via \eqref{eq:cf}.
Causal forests have several tuning parameters (e.g., minimum node size for individual trees), and
we choose those tuning parameters by cross-validation on the $R$-objective \eqref{eq:rlearner}, i.e.,
we train causal forests with different values of the tuning parameters, and choose the ones that make
out-of-bag estimates of the objective minimized in \eqref{eq:rlearner} as small as possible.

\begin{algorithm}[t]
\begin{center}
\begin{lstlisting}
Y.forest = regression_forest(X, Y, clusters = school.id)
Y.hat = predict(Y.forest)$predictions
W.forest = regression_forest(X, W, clusters = school.id)
W.hat = predict(W.forest)$predictions

cf.raw = causal_forest(X, Y, W,
                       Y.hat = Y.hat, W.hat = W.hat,
                       clusters = school.id)
varimp = variable_importance(cf.raw)
selected.idx = which(varimp > mean(varimp))

cf = causal_forest(X[,selected.idx], Y, W,
                   Y.hat = Y.hat, W.hat = W.hat,
                   clusters = school.id,
                   samples_per_cluster = 50,
                   tune.parameters = TRUE)
tau.hat = predict(cf)$predictions
\end{lstlisting}
\caption{Estimating treatment effects with causal forests.}
\label{alg:main}
\end{center}
\end{algorithm}

We provide an exact implementation of our treatment effect estimation strategy with causal
forests in Algorithm \ref{alg:main}. We train the \texttt{Y.forest} and \texttt{W.forest} using default
settings, as their predictions are simply used as inputs to the causal forest and default
parameter choices often perform reasonably well with random forests.\footnote{The
nuisance components \texttt{Y.hat} or \texttt{W.hat} need not be estimated by a regression
forest. We could also use other predictive methods (e.g., boosting with cross-fitting) or use oracle
values (e.g., the true randomization probabilities for \texttt{W.hat} in a randomized trial).
If we simply run the command \texttt{causal\char`_forest(X, Y, W)} without specifying
\texttt{Y.hat} or \texttt{W.hat}, then the software silently estimates \texttt{Y.hat} or \texttt{W.hat}
via regression forests.}
For our final causal forest, however, we deploy some tweaks for improved precision.
Motivated by \citet*{basu2018iterative}, we start by training a pilot random forest
on all the features, and then train a second forest on only those features that
saw a reasonable number of splits in the first step.\footnote{Given good estimates of 
\texttt{Y.hat} and \texttt{W.hat}, the construction \eqref{eq:cf} eliminates confounding
effects. Thus, we do not need to give the causal forest all features $X$ that may be confounders.
Rather, we can focus on features that we believe may be treatment modifiers; see
\citet*{zhao2017selective} for a further discussion.}
This enables the forest to make
more splits on the most important features in low-signal situations. Second, we increase
the \texttt{samples\char`_per\char`_cluster} parameter (called $k$ in Section \ref{sec:clust}) to
increase the number of samples used to grow each tree. Finally, the option
\texttt{tune.parameters = TRUE} has the forest cross-validate tuning parameters using
the $R$-objective rather than just setting defaults.

\section{Workshop Results}
\label{sec:results}

We now use our causal forest as trained in Algorithm \ref{alg:main} to explore the questions from Section \ref{sec:data}.

\subsection{The average treatment effect}
\label{sec:ate}
The first question asks about the overall effectiveness of the
intervention. The package \texttt{grf} has a built-in function for average treatment effect estimation,
based on a variant of augmented inverse-propensity weighting \citep*{robins1994estimation}.
With clusters, we compute an average treatment effect estimate \smash{$\htau$} and a standard
error estimate \smash{$\hsigma^2$} as follows:
\begin{equation}
\label{eq:atehat}
\begin{split}
&\htau_j = \frac{1}{n_j} \sum_{\cb{i : A_i = j}} \hGamma_i, \ \ \ \ 
\htau = \frac{1}{J} \sum_{j = 1}^J \htau_j, \ \ \ \
\hsigma^2 = \frac{1}{J(J-1)} \sum_{j = 1}^J \p{\htau_j - \htau}^2, \\
&\hGamma_i = \htau^{(-i)}\p{X_i} + \frac{W_i - \he^{(-i)}(X_i)}{\he^{(-i)}\p{X_i} \p{1 - \he^{(-i)}\p{X_i}}}
\p{Y_i - \hatm^{(-i)}(X_i) - \p{W_i - \he^{(-i)}\p{X_i}}\htau^{(-i)}\p{X_i}}.
\end{split}
\end{equation}
See Section 2.1 of \citet{farrell2015robust} for a discussion of estimators with this functional form,
and Section 2.4 of \citet*{athey2018approximate} for a recent literature review. The value of
cross-fitting is stressed in \citet{chernozhukov2017double}.
An application of this method suggests that the treatment had a large positive on average.

\begin{center}
\vbox{\rule{\textwidth}{0.1mm}
\begin{lstlisting}
ATE = average_treatment_effect(cf)
paste("95% CI for the ATE:", round(ATE[1], 3),
      "+/-", round(qnorm(0.975) * ATE[2], 3))
> "95% CI for the ATE: 0.247 +/- 0.04"
\end{lstlisting}
\vspace{1.5mm}
\rule{\textwidth}{0.1mm}}
\end{center}

\begin{figure}[h]
\begin{centering}
\includegraphics[width=0.45\textwidth]{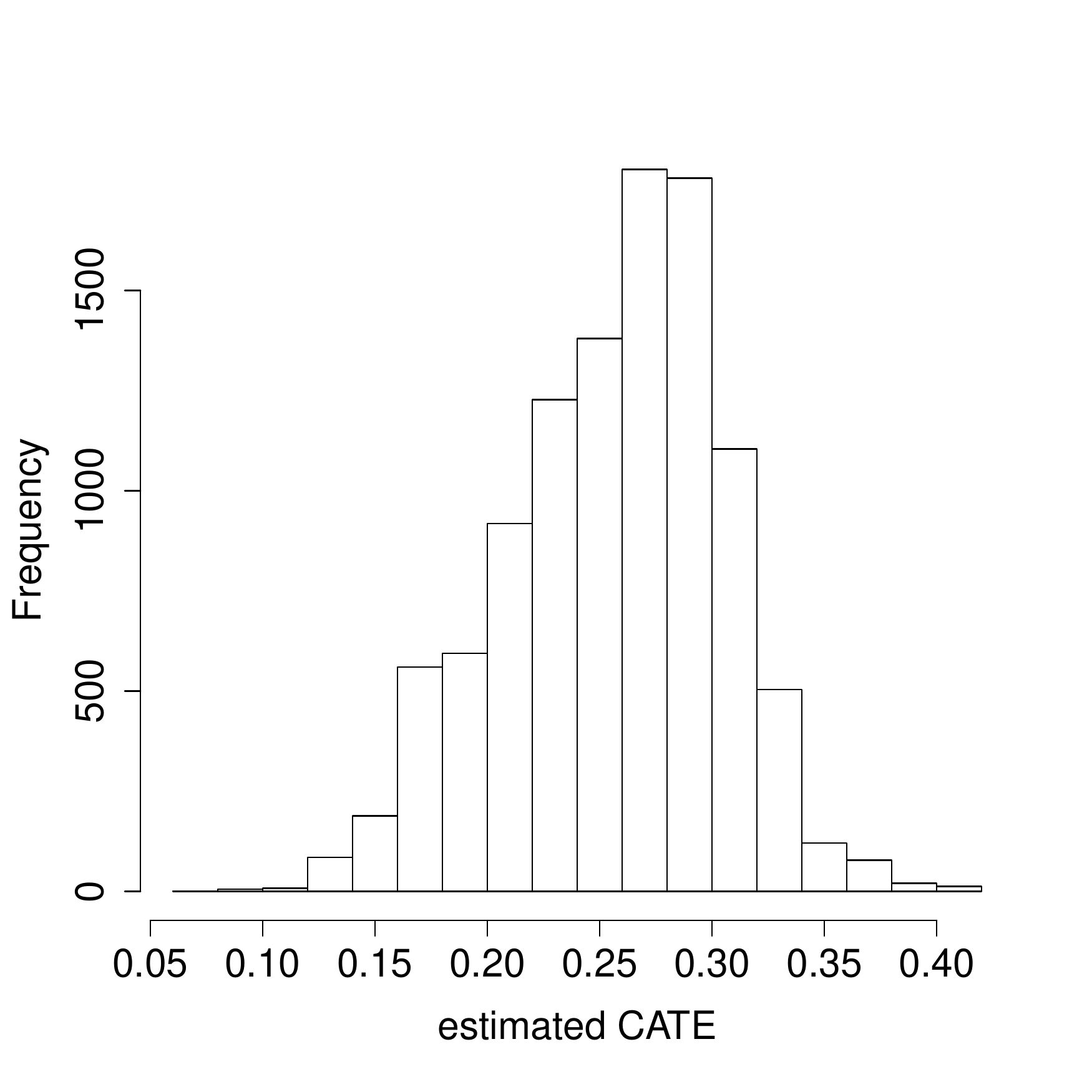}\
\caption{Histogram of out-of-bag CATE estimates from a causal forest
trained as in Algorithm \ref{alg:main}.}
\label{fig:tau_hist}
\end{centering}
\end{figure}

\subsection{Assessing treatment heterogeneity}
\label{sec:hte}
The next two questions pertain to treatment heterogeneity. Before addressing questions, however,
it is useful to ask whether the causal forest has succeeded in accurately estimating treatment heterogeneity.
As seen in Figure \ref{fig:tau_hist}, the causal forest CATE estimates obviously
exhibit variation; but this does not automatically imply that $\htau^{(-i)}(X_i)$ is a better
estimate of $\tau(X_i)$ than the overall average treatment effect estimate $\htau$ from
\eqref{eq:atehat}. Below, we seek an overall hypothesis test for whether heterogeneity in
$\htau^{(-i)}(X_i)$ is associated with heterogeneity in $\tau(X_i)$.

A first, simple approach to testing for heterogeneity involves grouping observations according
to whether their out-of-bag CATE estimates are above or below the median CATE estimate, and
then estimating average treatment effects in these two subgroups separately using the doubly
robust approach \eqref{eq:atehat}. This procedure is somewhat heuristic, as the ``high'' and ``low''
subgroups are not independent of the scores \smash{$\hGamma_i$} used to estimate the within-group
effects; however, the subgroup definition does not directly depend on the outcomes or treatments $(Y_i, W_i)$
themselves, and it appears that this approach can provide at least qualitative insights about the strength
of heterogeneity.

We also try a second test for heterogeneity, motivated by the
``best linear predictor'' method of \citet*{chernozhukov2018generic},
that seeks to fit the CATE as a linear function of the the out-of-bag causal forest estimates
$\htau^{(-i)}(X_i)$. Concretely, following \eqref{eq:robinson}, we create two synthetic predictors,
\smash{$C_i = \btau (W_i - \he^{(-i)}(X_i))$} and \smash{$D_i = (\htau^{(-i)}(X_i) - \btau) (W_i - \he^{(-i)}(X_i))$}
where $\btau$ is the average of the out-of-bag treatment effect estimates, and regress
\smash{$Y_i - \hatm^{(-i)}(X_i)$} against $C_i$ and $D_i$. Then, we can interpret
the coefficient of $D_i$ as a measure of the quality of the estimates of treatment heterogeneity,
while $C_i$ absorbs the average treatment effect. If the coefficient on $D_i$ is 1, then the treatment
heterogeneity estimates are well calibrated, while if the coefficient is $D_i$ significant and positive,
then at least we have evidence of a useful association between $\htau^{(-i)}(X_i)$ and $\tau(X_i)$.
More formally, one could use the $p$-value for the coefficient of $D_i$ to test the hypothesis that
the causal forest succeeded in finding heterogeneity; however, we caution that asymptotic results
justifying such inference are not presently available.

Below, we show output from running both analyses (note that all results are cluster-robust,
where each cluster gets the same weight). The overall picture appears somewhat mixed: Although
point estimates are consistent with the presence of heterogeneity, neither detection is significant.
Thus, at least if we insist on cluster-robust inference, any treatment heterogeneity that may be present
appears to be relatively weak, and causal forests do not identify subgroups with effects that obviously
stand out. We discuss the role of cluster-robustness further in Section \ref{sec:clust_eval}.

\begin{center}
\vbox{\rule{\textwidth}{0.1mm}
\begin{lstlisting}
# Compare regions with high and low estimated CATEs
high_effect = tau.hat > median(tau.hat)
ate.high = average_treatment_effect(cf, subset = high_effect)
ate.low = average_treatment_effect(cf, subset = !high_effect)
paste("95% CI for difference in ATE:",
      round(ate.high[1] - ate.low[1], 3), "+/-",
      round(qnorm(0.975) * sqrt(ate.high[2]^2 + ate.low[2]^2), 3))
> "95% CI for difference in ATE: 0.053 +/- 0.071"

# Run best linear predictor analysis
test_calibration(cf)
>                         Estimate Std. Error t value Pr(>|t|)    
> mean.prediction         1.007477   0.083463 12.0710   <2e-16 ***
> differential.prediction 0.321932   0.306738  1.0495    0.294    
\end{lstlisting}
\vspace{1.5mm}
\rule{\textwidth}{0.1mm}}
\end{center}

\begin{figure}
\begin{centering}
\begin{tabular}{cc}
\includegraphics[width=0.45\textwidth]{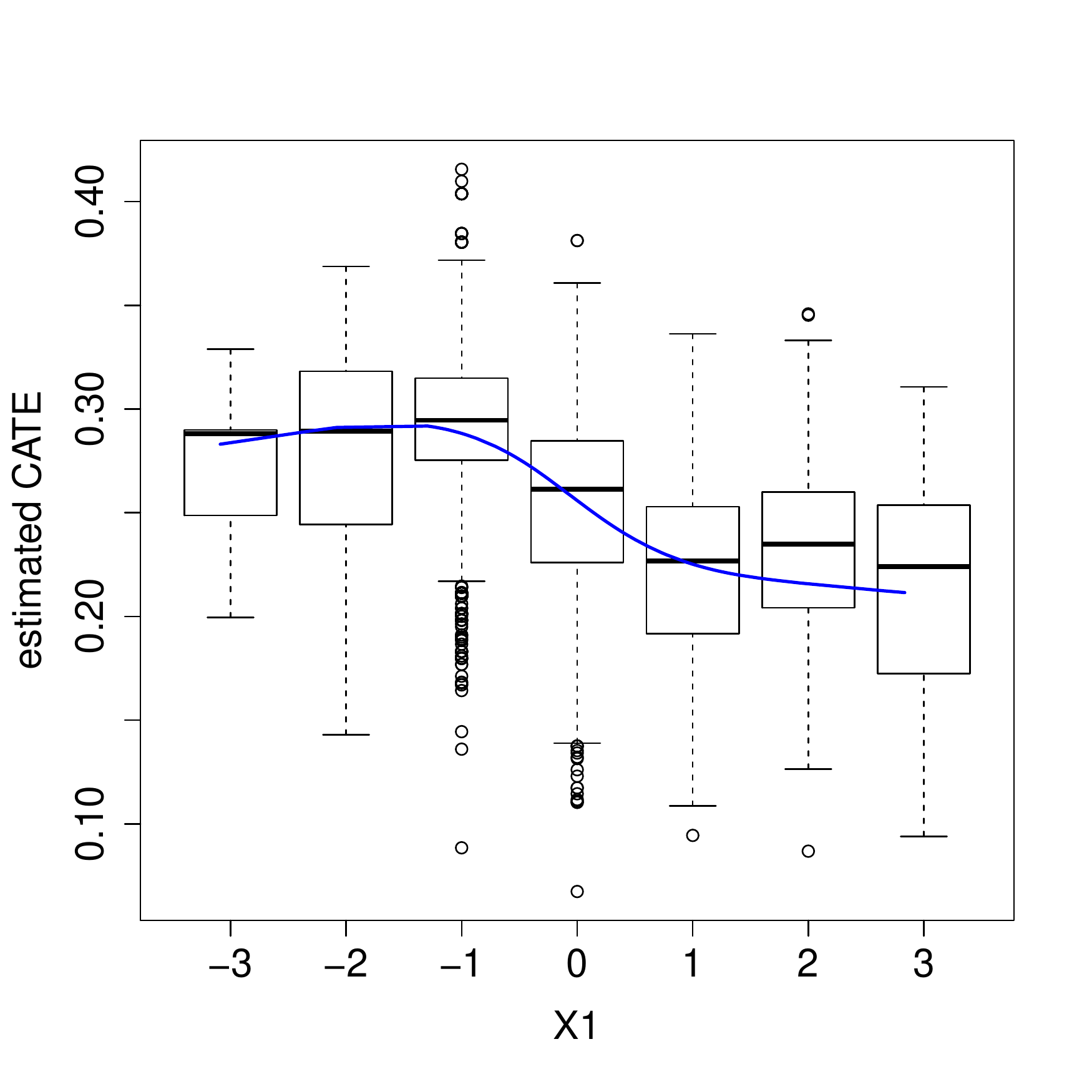} &
\includegraphics[width=0.45\textwidth]{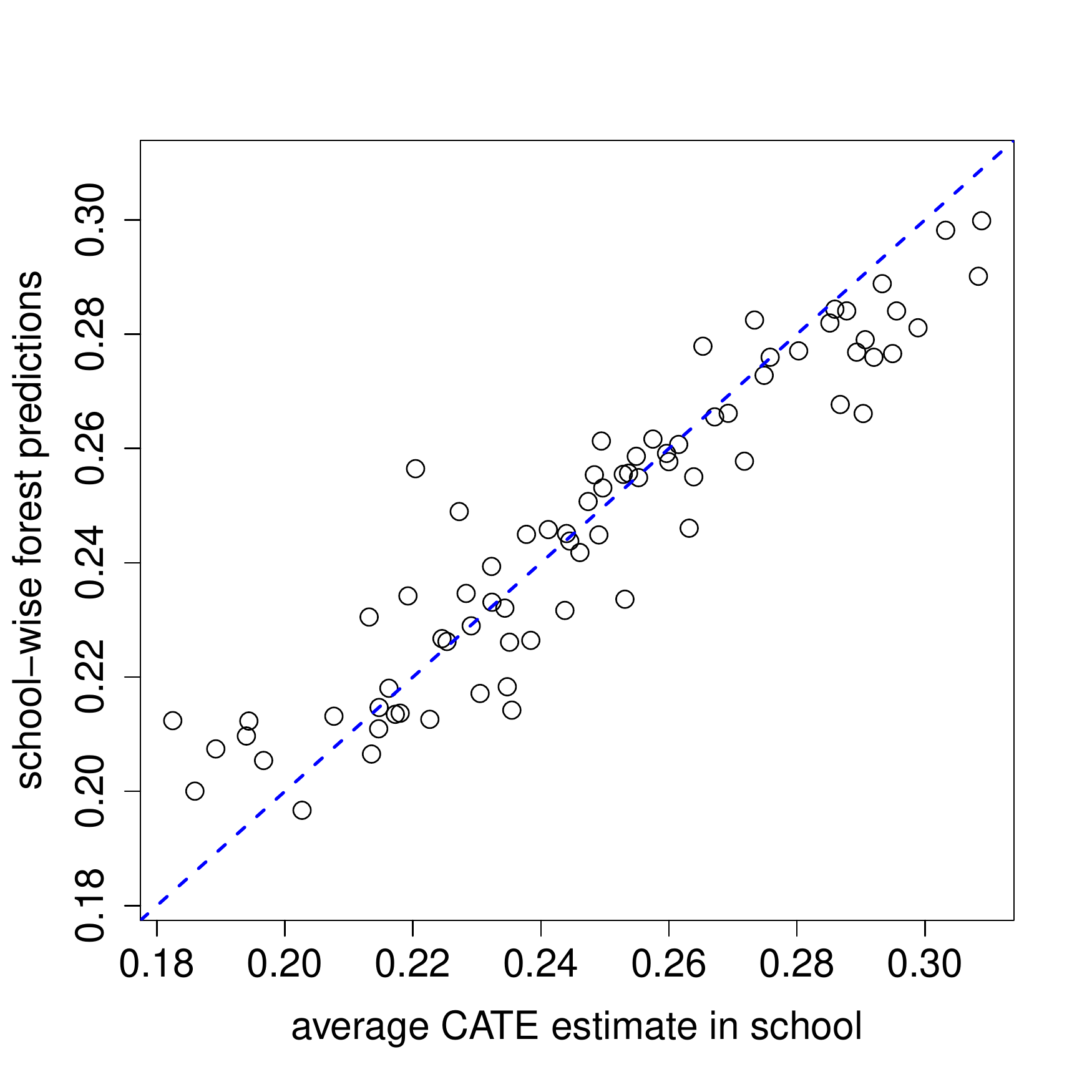} \\
{\bf (a)} variation with school-level mindset &
{\bf (b)} evaluating forest trained on $\htau_j$ from \eqref{eq:atehat} \\
\end{tabular}
\caption{Panel {\bf (a)} plots students' CATE estimates against
school-level mindset \texttt{X1}. Panel {\bf (b)} compares
estimates from a regression forest trained to predict the per-school doubly robust
treatment effect estimates $\htau_j$ from \eqref{eq:atehat} using school-level covariates,
to school-wise averages of the causal forest estimates $\htau^{(-i)}(X_i)$ trained as in
Algorithm \ref{alg:main}.}
\label{fig:tauhat}
\end{centering}
\end{figure}

\subsection{The effect of \texttt{X1} and \texttt{X2}}
\label{sec:x1x2}
Although our omnibus tests did not find strong evidence of treatment heterogeneity,
this does not mean there is no heterogeneity present. Researchers had pre-specified
interest in heterogeneity along two specific variables, namely pre-existing mindset (\texttt{X1})
and school-level achievement (\texttt{X2}), and it is plausible that a test for heterogeneity that
focuses on these two variables may have more power than the agnostic tests explored
above.

Both \texttt{X1} and \texttt{X2} are school-level variables, so we here design tests based on the
per-school doubly robust treatment effect estimates \smash{$\htau_j$} computed
in \eqref{eq:atehat}. As seen below, this more targeted analysis uncovers notable
heterogeneity along \texttt{X1}, i.e., schools with larger values of \texttt{X1} appear to experience
larger effects than schools with smaller values of \texttt{X1}. Conversely, we do not see
much heterogeneity along \texttt{X2}, whether we divide schools into 2 subgroups (to
test the monotone hypothesis) or into 3 subgroups (to test the goldilocks hypothesis).

Although the $p$-value for heterogeneity along \texttt{X1} is not small enough to withstand a Bonferroni
test, it seems reasonable to take the detection along \texttt{X1} seriously because heterogeneity
along \texttt{X1} was one of two pre-specified hypotheses.
Interestingly, we also note that \texttt{X1} was the most important variable in the causal forest:
The final causal forest was trained on 9 ``selected'' variables, and spent 24\% of its splits on \texttt{X1}
with splits weighted by depth (as in the function \texttt{variable\char`_importance}). The left panel of
Figure \ref{fig:tauhat} plots the relationship between \texttt{X1} and $\htau^{(-i)}(X_i)$.

\begin{center}
\vbox{\rule{\textwidth}{0.1mm}
\begin{lstlisting}
dr.score = tau.hat + W / cf$W.hat *
  (Y - cf$Y.hat - (1 - cf$W.hat) * tau.hat) -
  (1 - W) / (1 - cf$W.hat) * (Y - cf$Y.hat + cf$W.hat * tau.hat)
school.score = t(school.mat) %*% dr.score / school.size

school.X1 = t(school.mat) %*% X$X1 / school.size
high.X1 = school.X1 > median(school.X1)
t.test(school.score[high.X1], school.score[!high.X1])
> t = -3.0205, df = 72.087, p-value = 0.00349
> 95 percent confidence interval: -0.1937 -0.0397

school.X2 = (t(school.mat) %*% X$X2) / school.size
high.X2 = school.X2 > median(school.X2)
t.test(school.score[high.X2], school.score[!high.X2])
> t = 1.043, df = 72.431, p-value = 0.3004
> 95 percent confidence interval: -0.0386 0.1234

school.X2.levels = cut(school.X2,
  breaks = c(-Inf, quantile(school.X2, c(1/3, 2/3)), Inf))
summary(aov(school.score ~ school.X2.levels))
>                  Df Sum Sq Mean Sq F value Pr(>F)
> school.X2.levels  2  0.085 0.04249   1.365  0.262
> Residuals        73  2.272 0.03112               
\end{lstlisting}
\vspace{1.5mm}
\rule{\textwidth}{0.1mm}}
\end{center}

\subsection{Looking for school-level heterogeneity}
\label{sec:school}
Our omnibus test for heterogeneity from Section \ref{sec:hte} produced
mixed results; however, when we zoomed in on the pre-specified covariates
\texttt{X1} and \texttt{X2} in Section \ref{sec:x1x2}, we uncovered interesting results.
Noticing that both \texttt{X1} and \texttt{X2} are school-level (as opposed to student-level) covariates,
it is natural to ask whether an analysis that only focuses only on school-level effects
may have had more power than our original analysis following Algorithm \ref{alg:main}.

Here, we examine this question by fitting models to the school-level estimates $\htau_j$
from \eqref{eq:atehat} using only school level covariates. We considered both an analysis
using a regression forest, as well as classical linear regression modeling. Both methods,
however, result in conclusions that are in line with the ones obtained above. The strength
of the heterogeneity found by the regression forest trained on the $\htau_j$ as measured by
the ``calibration test'' is comparable to the strength of the heterogeneity found by our original
causal forest; moreover, as seen in the right panel of Figure \ref{fig:tauhat}, the predictions
made by this regression forest are closely aligned with school-wise averaged predictions from
the original causal forest. Meanwhile, a basic linear regression analysis uncovers a borderline
amount of effect modification along \texttt{X1} and nothing else stands out.

The overall picture is that, by looking at the predictor \texttt{X1} alone, we can find credible
effect modification that is correlated negatively with \texttt{X1}. However, there does not
appear to be strong enough heterogeneity for us to be able to accurately fit a more complex model for $\tau(\cdot)$:
Even a linear model for effect modification starts to suffer from low signal, and it is not quite
clear whether \texttt{X1} is an effect modifier after we control for the other school-level covariates.

\begin{center}
\vbox{\rule{\textwidth}{0.1mm}
\begin{lstlisting}
# Regression forest analysis
school.forest = regression_forest(school.X, school.score)
school.pred = predict(school.forest)$predictions
test_calibration(school.forest)
>                         Estimate Std. Error t value Pr(>|t|)    
> mean.prediction          0.998765   0.083454 11.9679   <2e-16 ***
> differential.prediction  0.619299   0.706514  0.8766   0.3836    

# Ordinary least-squares analysis
coeftest(lm(school.score ~ school.X), vcov = vcovHC)
>               Estimate Std. Error t value Pr(>|t|)   
> (Intercept)  0.2434703  0.0770302  3.1607 0.002377 **
> X1          -0.0493032  0.0291403 -1.6919 0.095377 . 
> X2           0.0143625  0.0340139  0.4223 0.674211   
> X3           0.0092693  0.0264267  0.3508 0.726888   
> X4           0.0248985  0.0258527  0.9631 0.339019   
> X5          -0.0336325  0.0265401 -1.2672 0.209525   
> XC.1        -0.0024447  0.0928801 -0.0263 0.979081   
> XC.2         0.0826898  0.1052411  0.7857 0.434845   
> XC.3        -0.1376920  0.0876108 -1.5716 0.120818   
> XC.4         0.0408624  0.0820938  0.4978 0.620313   
\end{lstlisting}
\vspace{1.5mm}
\rule{\textwidth}{0.1mm}}
\end{center}

\section{Post-workshop analysis}
\label{sec:post}

Two notable differences between the causal forest analysis used here and a more
direct machine-learning-based analysis were our use of cluster-robust methods, and of
orthogonalization for robustness to confounding as in \eqref{eq:cf}. To understand the
value of these features, we revisit some analyses from Section \ref{sec:results} without
them.

\subsection{The value of clustering}
\label{sec:clust_eval}

\begin{figure}
\begin{centering}
\begin{tabular}{cc}
\includegraphics[width=0.45\textwidth]{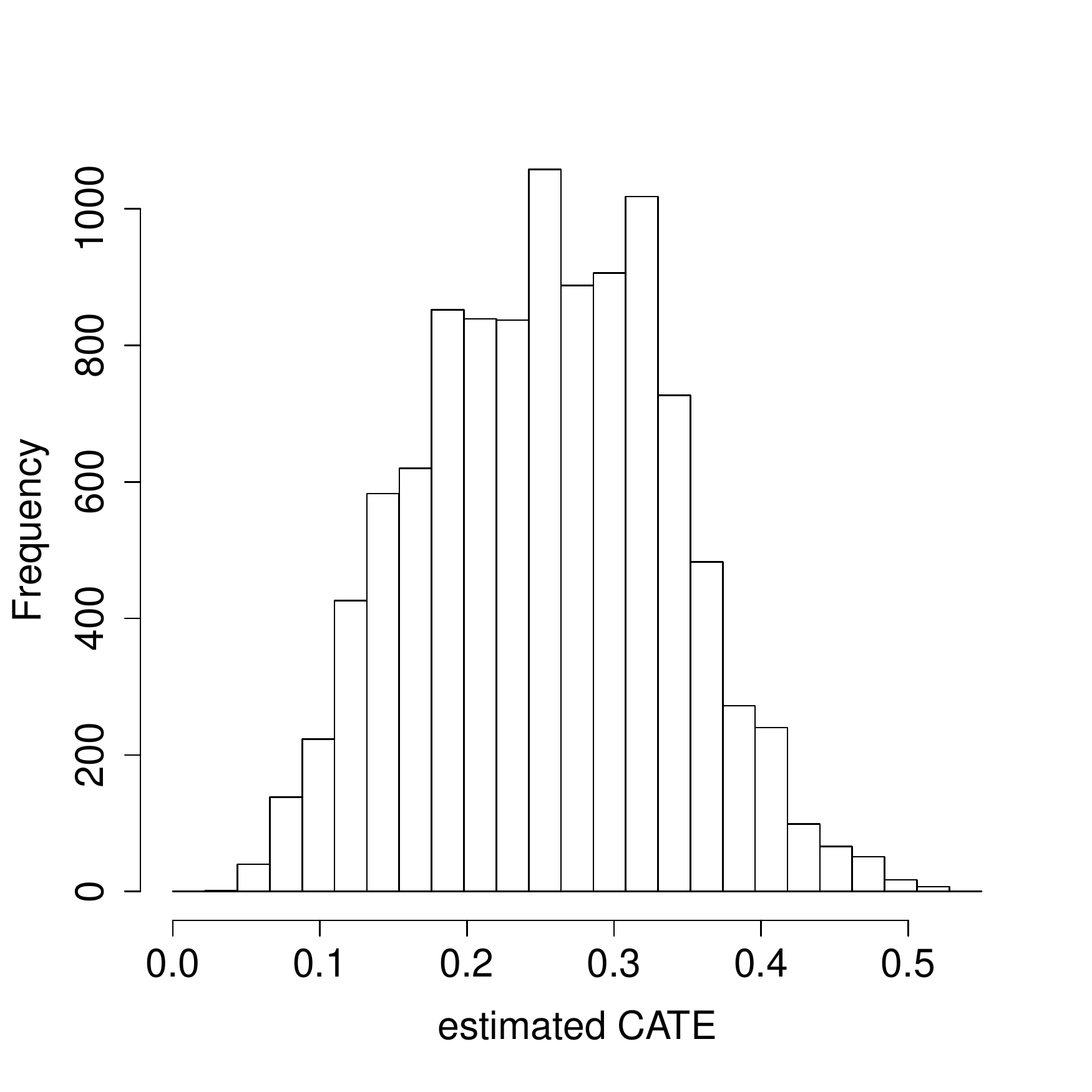} &
\includegraphics[width=0.45\textwidth]{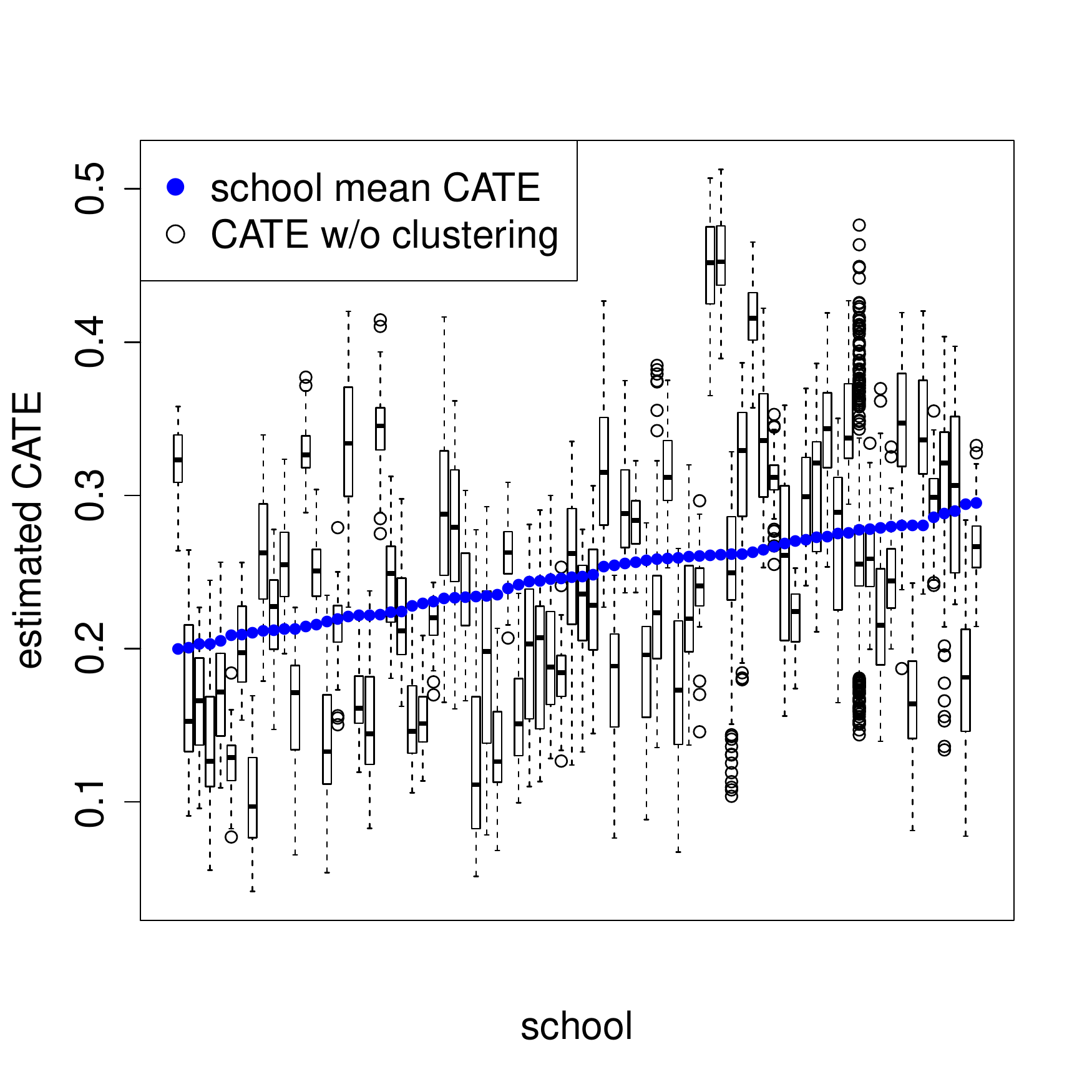} \\
{\bf (a)} histogram of CATE estimates w/o clustering &
{\bf (b)} per-school CATE estimates w/o clustering
\end{tabular}
\caption{Panel {\bf (a)} is a histogram of CATE estimates \smash{$\htau^{(-i)}(X_i)$} trained
using a causal forest that does not account for school-level clustering.
Panel {\bf (b)} compares per-student predictions $\htau^{(-i)}(X_i)$ from a non-cluster-robust
causal forest to per-school mean treatment effect
predictions from a forest trained on per-school responses as in Section \ref{sec:school}.} 
\label{fig:comparison}
\end{centering}
\end{figure}

If we train a causal forest on students without clustering by school, we obtain markedly
different results from before: The confidence interval for the average treatment effect is now roughly half as
long as before, and there appears to be unambiguously detectable heterogeneity
according to the \texttt{test\char`_calibration} function. Moreover, as seen in the
left panel of Figure \ref{fig:comparison}, the CATE estimates \smash{$\htau^{(-i)}(X_i)$}
obtained without clustering are much more dispersed than those obtained with clustering
(see Figure \ref{fig:tau_hist}): The sample variance of the \smash{$\htau^{(-i)}(X_i)$} increases by
a factor 5.82 without clustering.

It appears that these strong detections without clustering are explained by
excess optimism from ignoring variation due to idiosyncratic school-specific effects, rather than from a true gain
in power from using a version of causal forests without clustering.
The right panel of Figure \ref{fig:comparison} shows per-school estimates of \smash{$\htau^{(-i)}(X_i)$} from
the non-cluster-robust causal forest, and compares them to predictions for the mean CATE in the school
obtained in a way that is cluster-robust. The differences are striking: For example, the left-most school
in the right panel of Figure \ref{fig:comparison} has non-cluster-robust \smash{$\htau^{(-i)}(X_i)$}
estimates that vary from 0.26 to 0.36, whereas the cluster-robust estimate of its mean CATE was roughly 0.2.
A simple explanation for how this could come about is that students in the
school happened to have unusually high treatment effects, and that the non-cluster-robust forest was able
to overfit to this school-level effect because it does not account for potential correlations
between different students in the same school.

To gain deeper insights into the behavior of non-cluster robust forests,
we tried a 5-fold version of this algorithm where the forests themselves are not
cluster-robust, but the estimation folds are cluster aligned. Specifically, we split
the schools into 5 folds; then, for each fold, we fit a causal forest without clustering on observations
belonging to schools in the 4/5 other folds, and made CATE estimates on the held out fold.
Finally, re-running a best linear prediction test on out-of-fold predictions as in the \texttt{test\char`_calibration} function,
we found at best tenuous evidence for the presence of heterogeneity (in fact, the resulting $t$-statistic
for heterogeneity, 0.058, was weaker than the one in Section \ref{sec:hte}).
In other words, if we use evaluation methods that are robust to clustering, then the apparent
gains from non-cluster-robust forests wash away.

Thus, it appears that different schools have very different values of $\htau_j$; however, most of the school-wise
effects appear to be idiosyncratic, and cannot be explained using covariates. In order to gain insights that generalize
to new schools we need to cluster by school; and, once we do so, much of the apparent heterogeneity between
schools ends up looking like noise.

\begin{center}
\vbox{\rule{\textwidth}{0.1mm}
\begin{lstlisting}
cf.noclust = causal_forest(X[,selected.idx], Y, W,
                           Y.hat = Y.hat, W.hat = W.hat,
                           tune.parameters = TRUE)
ATE.noclust = average_treatment_effect(cf.noclust)
paste("95% CI for the ATE:", round(ATE.noprop[1], 3),
      "+/-", round(qnorm(0.975) * ATE.noprop[2], 3))
> "95% CI for the ATE: 0.253 +/- 0.022"

test_calibration(cf.noclust)
>                         Estimate Std. Error t value  Pr(>|t|)    
> mean.prediction         1.003796   0.044779 22.4164 < 2.2e-16 ***
> differential.prediction 0.634163   0.132700  4.7789 1.786e-06 ***
\end{lstlisting}
\vspace{1.5mm}
\rule{\textwidth}{0.1mm}}
\end{center}

\begin{figure}
\begin{centering}
\includegraphics[width=0.45\textwidth]{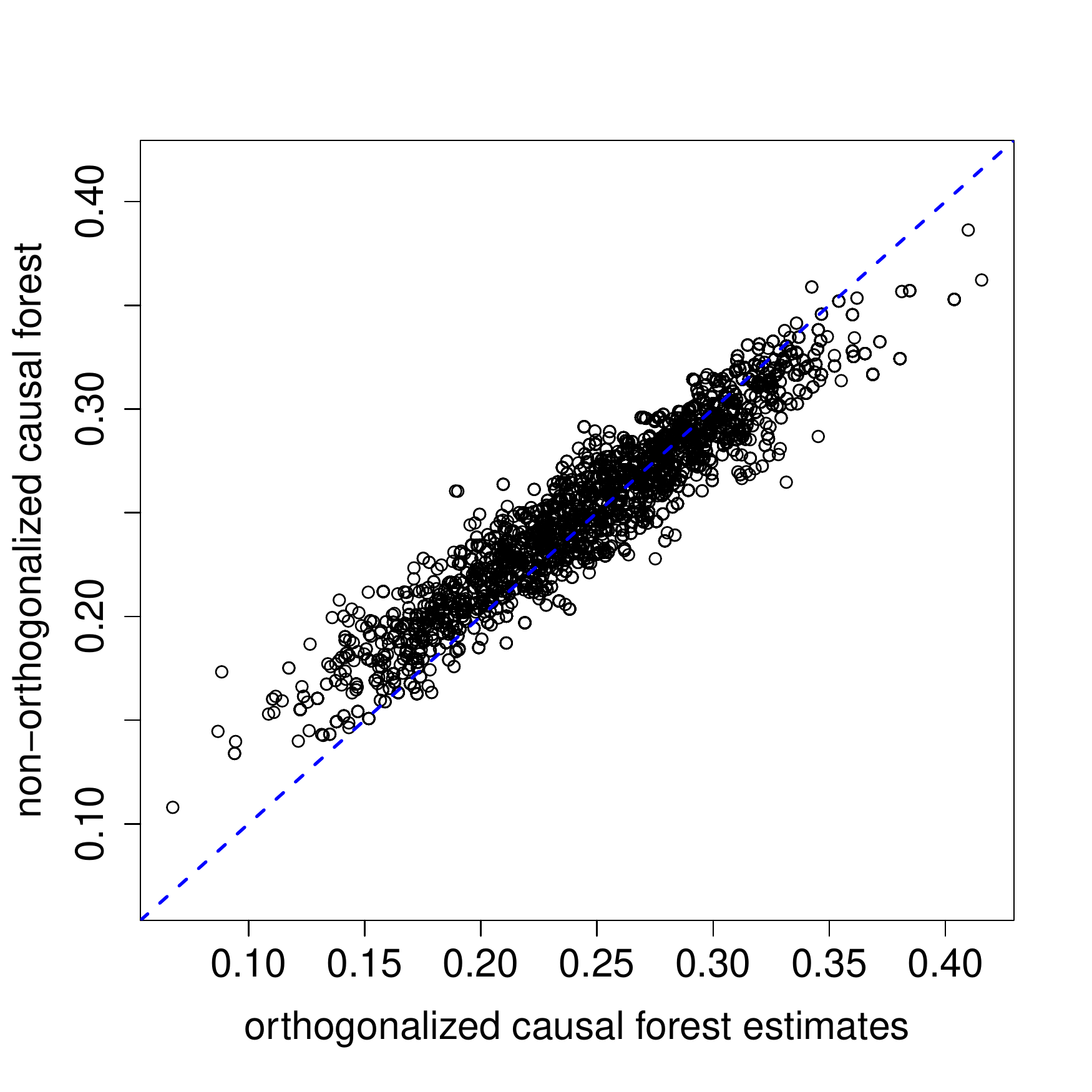}\
\caption{Comparison of  estimates from a forest trained with a
trivial propensity model \smash{$\he(X_i) = \overline{W} = n^{-1} \sum_{i = 1}^n W_i$}
to predictions from the forest trained as in Algorithm \ref{alg:main}.}
\label{fig:noprop}
\end{centering}
\end{figure}

\subsection{The value of orthogonalization}
\label{sec:orthog_eval}

In this dataset, orthogonalization appears to be less important than clustering.
If we train a causal forests without estimating the propensity score or, more
specifically, using the trivial propensity model
\smash{$\he(X_i) = \overline{W} = n^{-1} \sum_{i = 1}^n W_i$}, we uncover essentially
the same average treatment effect estimate as with orthogonalization. Moreover, as
shown in Figure \ref{fig:noprop}, the causal forests trained
with or without orthogonalization yield essentially the same CATE estimates \smash{$\htau^{(-i)}(X_i)$}.

One reason for this phenomenon may be that, here, the most important
confounders are also important for predicting $Y$: In Algorithm \ref{alg:main}, the
most important predictor for both the $W$- and $Y$-forests is \texttt{S3}, with
22\% of splits and 70\% of splits respectively (both weighted by depth as in the
\texttt{variable\char`_importance} function). Meanwhile, as argued in
\citet*{belloni2014inference}, orthogonalization is often most important when
there are some features that are highly predictive of treatment propensities but
not very predictive of $Y$. Thus, it is possible that the non-orthogonalized forest
does well here because we were lucky, and there were no confounders that
only had a strong effect the propensity model.

To explore this hypothesis, we present a synthetic example where some
variables have stronger effects on $W$ than on $Y$ and see that, as expected,
orthogonalization is now important. There is clearly no treatment effect, yet the
non-orthogonalized forest appears to find a non-zero effect.

\begin{center}
\vbox{\rule{\textwidth}{0.1mm}
\begin{lstlisting}
cf.noprop = causal_forest(X[,selected.idx], Y, W,
                          Y.hat = Y.hat, W.hat = mean(W),
                          tune.parameters = TRUE,
                          samples_per_cluster = 50,
                          clusters = school.id)
ATE.noprop = average_treatment_effect(cf.noprop)
paste("95% CI for the ATE:", round(ATE.noprop[1], 3),
      "+/-", round(qnorm(0.975) * ATE.noprop[2], 3))
> "95% CI for the ATE: 0.253 +/- 0.04"

n.synth = 1000; p.synth = 10
X.synth = matrix(rnorm(n.synth * p.synth), n.synth, p.synth)
W.synth = rbinom(n.synth, 1, 1 / (1 + exp(-X.synth[,1])))
Y.synth = 2 * rowMeans(X.synth[,1:6]) + rnorm(n.synth)
Y.forest.synth = regression_forest(X.synth, Y.synth)
Y.hat.synth = predict(Y.forest.synth)$predictions
W.forest.synth = regression_forest(X.synth, W.synth)
W.hat.synth = predict(W.forest.synth)$predictions

cf.synth = causal_forest(X.synth, Y.synth, W.synth,
            Y.hat = Y.hat.synth, W.hat = W.hat.synth)
ATE.synth = average_treatment_effect(cf.synth)
paste("95% CI for the ATE:", round(ATE.synth[1], 3),
      "+/-", round(qnorm(0.975) * ATE.synth[2], 3))
> "95% CI for the ATE: 0.125 +/- 0.151"

cf.synth.noprop = causal_forest(X.synth, Y.synth, W.synth,
            Y.hat = Y.hat.synth, W.hat = mean(W.synth))
ATE.synth.noprop = average_treatment_effect(cf.synth.noprop)
paste("95% CI for the ATE:", round(ATE.synth.noprop[1], 3),
      "+/-", round(qnorm(0.975) * ATE.synth.noprop[2], 3))
> "95% CI for the ATE: 0.220 +/- 0.142"
\end{lstlisting}
\vspace{1.5mm}
\rule{\textwidth}{0.1mm}}
\end{center}

\noindent

\section{Discussion}

We applied causal forests to study treatment heterogeneity on a dataset derived from
the National Study of Learning Mindsets. Two challenges in this setting involved
an observational study design with unknown treatment propensities, and clustering of
outcomes at the school level. Causal forests allow for an
algorithmic specification that addresses both challenges.
Of these two challenges, school-level clustering had a dramatic effect on our analysis. If we properly
account for the clustering, we find hints of the presence of treatment heterogeneity
(Section \ref{sec:x1x2}), but accurate non-parametric estimation of $\tau(x)$ is difficult (Section \ref{sec:hte}).
In contrast, an analysis that ignores clusters claims to find very strong heterogeneity in $\tau(x)$
that can accurately be estimated (Section \ref{sec:clust_eval}).

This result highlights the need for a deeper discussion of the how to work with
clustered observations when modeling treatment heterogeneity. The traditional approach
is to capture cluster effects via ``fixed effect'' or ``random effect'' models of the form
\begin{equation}
\label{eq:FE}
Y_i = m(X_i) + W_i \tau(X_i) + \beta_{A_i} + W_i \gamma_{A_i} + \varepsilon_i,
\end{equation}
where $A_i \in \cb{1, \, ..., \, J}$ denotes the cluster membership of the $i$-th sample
whereas $\beta_j$ and $\gamma_j$ denote per-cluster offsets on the main effect and
treatment effect respectively, and the nomenclature around fixed or random effects 
reflects modeling choices for $\beta$ and $\gamma$ \citep{wooldridge2010econometric}.
In a non-parametric setting, however, assuming that clusters have an additive effect
on $Y_i$ seems rather restrictive. The approach we took in this note can be interpreted
as fitting a functional random effects model
\begin{equation}
\label{eq:FRE}
Y_i = m_{A_i}(X_i) + W_i \tau_{A_i}(X_i) + \varepsilon_i, \ \ \tau(x) = \EE{\tau_j(x)},
\end{equation}
where each cluster has its own main and treatment effect function, and the expectation
above is defined with respect to the distribution of per-cluster treatment effect functions.
It would be of considerable interest to develop a better understanding of the pros and
cons of different approaches to heterogeneous treatment effect estimation on clustered
data.

\bibliographystyle{plainnat-abbrev}
\bibliography{references}

\begin{thebibliography}{35}
\providecommand{\natexlab}[1]{#1}
\providecommand{\url}[1]{\texttt{#1}}
\expandafter\ifx\csname urlstyle\endcsname\relax
  \providecommand{\doi}[1]{doi: #1}\else
  \providecommand{\doi}{doi: \begingroup \urlstyle{rm}\Url}\fi

\bibitem[Abadie et~al.(2017)Abadie, Athey, Imbens, and
  Wooldridge]{abadie2017should}
A.~Abadie, S.~Athey, G.~Imbens, and J.~Wooldridge.
\newblock When should you adjust standard errors for clustering?
\newblock \emph{arXiv preprint arXiv:1710.02926}, 2017.

\bibitem[Angrist and Pischke(2008)]{angrist2008mostly}
J.~D. Angrist and J.-S. Pischke.
\newblock \emph{Mostly harmless econometrics: An empiricist's companion}.
\newblock Princeton university press, 2008.

\bibitem[Athey and Imbens(2016)]{athey2015machine}
S.~Athey and G.~Imbens.
\newblock Recursive partitioning for heterogeneous causal effects.
\newblock \emph{Proceedings of the National Academy of Sciences}, 113\penalty0
  (27):\penalty0 7353--7360, 2016.

\bibitem[Athey et~al.(2018)Athey, Imbens, and Wager]{athey2018approximate}
S.~Athey, G.~W. Imbens, and S.~Wager.
\newblock Approximate residual balancing: debiased inference of average
  treatment effects in high dimensions.
\newblock \emph{Journal of the Royal Statistical Society: Series B (Statistical
  Methodology)}, 80\penalty0 (4):\penalty0 597--623, 2018.

\bibitem[Athey et~al.(2019)Athey, Tibshirani, and Wager]{athey2018generalized}
S.~Athey, J.~Tibshirani, and S.~Wager.
\newblock Generalized random forests.
\newblock \emph{The Annals of Statistics}, 47\penalty0 (2):\penalty0
  1148--1178, 2019.

\bibitem[Basu et~al.(2018)Basu, Kumbier, Brown, and Yu]{basu2018iterative}
S.~Basu, K.~Kumbier, J.~B. Brown, and B.~Yu.
\newblock Iterative random forests to discover predictive and stable high-order
  interactions.
\newblock \emph{Proceedings of the National Academy of Sciences}, page
  201711236, 2018.

\bibitem[Belloni et~al.(2014)Belloni, Chernozhukov, and
  Hansen]{belloni2014inference}
A.~Belloni, V.~Chernozhukov, and C.~Hansen.
\newblock Inference on treatment effects after selection among high-dimensional
  controls.
\newblock \emph{The Review of Economic Studies}, 81\penalty0 (2):\penalty0
  608--650, 2014.

\bibitem[Biau and Scornet(2016)]{biau2016random}
G.~Biau and E.~Scornet.
\newblock A random forest guided tour.
\newblock \emph{Test}, 25\penalty0 (2):\penalty0 197--227, 2016.

\bibitem[Breiman(2001)]{breiman2001random}
L.~Breiman.
\newblock Random forests.
\newblock \emph{Machine Learning}, 45\penalty0 (1):\penalty0 5--32, 2001.

\bibitem[Chernozhukov et~al.(2018{\natexlab{a}})Chernozhukov, Chetverikov,
  Demirer, Duflo, Hansen, Newey, and Robins]{chernozhukov2017double}
V.~Chernozhukov, D.~Chetverikov, M.~Demirer, E.~Duflo, C.~Hansen, W.~Newey, and
  J.~Robins.
\newblock Double/debiased machine learning for treatment and structural
  parameters.
\newblock \emph{The Econometrics Journal}, 21\penalty0 (1):\penalty0 C1--C68,
  2018{\natexlab{a}}.

\bibitem[Chernozhukov et~al.(2018{\natexlab{b}})Chernozhukov, Demirer, Duflo,
  and Fernandez-Val]{chernozhukov2018generic}
V.~Chernozhukov, M.~Demirer, E.~Duflo, and I.~Fernandez-Val.
\newblock Generic machine learning inference on heterogenous treatment effects
  in randomized experiments.
\newblock Technical report, National Bureau of Economic Research,
  2018{\natexlab{b}}.

\bibitem[Ding et~al.(2016)Ding, Feller, and Miratrix]{ding2016randomization}
P.~Ding, A.~Feller, and L.~Miratrix.
\newblock Randomization inference for treatment effect variation.
\newblock \emph{Journal of the Royal Statistical Society: Series B (Statistical
  Methodology)}, 78\penalty0 (3):\penalty0 655--671, 2016.

\bibitem[Dorie et~al.(2017)Dorie, Hill, Shalit, Scott, and
  Cervone]{dorie2017automated}
V.~Dorie, J.~Hill, U.~Shalit, M.~Scott, and D.~Cervone.
\newblock Automated versus do-it-yourself methods for causal inference: Lessons
  learned from a data analysis competition.
\newblock \emph{arXiv preprint arXiv:1707.02641}, 2017.

\bibitem[Farrell(2015)]{farrell2015robust}
M.~H. Farrell.
\newblock Robust inference on average treatment effects with possibly more
  covariates than observations.
\newblock \emph{Journal of Econometrics}, 189\penalty0 (1):\penalty0 1--23,
  2015.

\bibitem[Gopalan and Tipton(2018)]{gopalan2018national}
M.~Gopalan and E.~Tipton.
\newblock Is the national study of learning mindsets nationally-representative?
\newblock \emph{PsyArXiv. November}, 3, 2018.

\bibitem[Hahn et~al.(2017)Hahn, Murray, and Carvalho]{hahn2017bayesian}
P.~R. Hahn, J.~S. Murray, and C.~Carvalho.
\newblock Bayesian regression tree models for causal inference: regularization,
  confounding, and heterogeneous effects.
\newblock \emph{arXiv preprint arXiv:1706.09523}, 2017.

\bibitem[Hill(2011)]{hill2011bayesian}
J.~L. Hill.
\newblock Bayesian nonparametric modeling for causal inference.
\newblock \emph{Journal of Computational and Graphical Statistics}, 20\penalty0
  (1), 2011.

\bibitem[Imai and Ratkovic(2013)]{imai2013estimating}
K.~Imai and M.~Ratkovic.
\newblock Estimating treatment effect heterogeneity in randomized program
  evaluation.
\newblock \emph{The Annals of Applied Statistics}, 7\penalty0 (1):\penalty0
  443--470, 2013.

\bibitem[Imbens and Rubin(2015)]{imbens2015causal}
G.~W. Imbens and D.~B. Rubin.
\newblock \emph{Causal Inference in Statistics, Social, and Biomedical
  Sciences}.
\newblock Cambridge University Press, 2015.

\bibitem[K{\"u}nzel et~al.(2017)K{\"u}nzel, Sekhon, Bickel, and
  Yu]{kunzel2017meta}
S.~K{\"u}nzel, J.~Sekhon, P.~Bickel, and B.~Yu.
\newblock Meta-learners for estimating heterogeneous treatment effects using
  machine learning.
\newblock \emph{arXiv preprint arXiv:1706.03461}, 2017.

\bibitem[Luedtke and van~der Laan(2016)]{luedtke2016super}
A.~R. Luedtke and M.~J. van~der Laan.
\newblock Super-learning of an optimal dynamic treatment rule.
\newblock \emph{The International Journal of Biostatistics}, 12\penalty0
  (1):\penalty0 305--332, 2016.

\bibitem[Nie and Wager(2017)]{nie2017learning}
X.~Nie and S.~Wager.
\newblock Quasi-oracle estimation of heterogeneous treatment effects.
\newblock \emph{arXiv preprint arXiv:1712.04912}, 2017.

\bibitem[Nosek et~al.(2015)]{nosek2015promoting}
B.~A. Nosek et~al.
\newblock Promoting an open research culture.
\newblock \emph{Science}, 348\penalty0 (6242):\penalty0 1422--1425, 2015.

\bibitem[{R Core Team}(2017)]{R}
{R Core Team}.
\newblock \emph{R: A Language and Environment for Statistical Computing}.
\newblock R Foundation for Statistical Computing, Vienna, Austria, 2017.
\newblock URL \url{https://www.R-project.org/}.

\bibitem[Robins et~al.(1994)Robins, Rotnitzky, and Zhao]{robins1994estimation}
J.~M. Robins, A.~Rotnitzky, and L.~P. Zhao.
\newblock Estimation of regression coefficients when some regressors are not
  always observed.
\newblock \emph{Journal of the American Statistical Association}, 89\penalty0
  (427):\penalty0 846--866, 1994.

\bibitem[Robinson(1988)]{robinson1988root}
P.~M. Robinson.
\newblock Root-n-consistent semiparametric regression.
\newblock \emph{Econometrica}, pages 931--954, 1988.

\bibitem[Rosenbaum(2002)]{rosenbaum2002observational}
P.~R. Rosenbaum.
\newblock \emph{Observational Studies}.
\newblock Springer, 2002.

\bibitem[Rosenbaum and Rubin(1983)]{rosenbaum1983central}
P.~R. Rosenbaum and D.~B. Rubin.
\newblock The central role of the propensity score in observational studies for
  causal effects.
\newblock \emph{Biometrika}, 70\penalty0 (1):\penalty0 41--55, 1983.

\bibitem[Shalit et~al.(2017)Shalit, Johansson, and
  Sontag]{shalit2017estimating}
U.~Shalit, F.~D. Johansson, and D.~Sontag.
\newblock Estimating individual treatment effect: generalization bounds and
  algorithms.
\newblock In \emph{ICML}, pages 3076--3085, 2017.

\bibitem[Su et~al.(2009)Su, Tsai, Wang, Nickerson, and Li]{su2009subgroup}
X.~Su, C.-L. Tsai, H.~Wang, D.~M. Nickerson, and B.~Li.
\newblock Subgroup analysis via recursive partitioning.
\newblock \emph{The Journal of Machine Learning Research}, 10:\penalty0
  141--158, 2009.

\bibitem[Tibshirani et~al.(2018)Tibshirani, Athey, Friedberg, Hadad, Miner,
  Wager, and Wright]{grf}
J.~Tibshirani, S.~Athey, R.~Friedberg, V.~Hadad, L.~Miner, S.~Wager, and
  M.~Wright.
\newblock \emph{grf: Generalized Random Forests (Beta)}, 2018.
\newblock URL \url{https://github.com/grf-labs/grf}.
\newblock R package version 0.10.2.

\bibitem[Wager and Athey(2018)]{wager2017estimation}
S.~Wager and S.~Athey.
\newblock Estimation and inference of heterogeneous treatment effects using
  random forests.
\newblock \emph{Journal of the American Statistical Association}, 113\penalty0
  (523):\penalty0 1228--1242, 2018.

\bibitem[Wooldridge(2010)]{wooldridge2010econometric}
J.~M. Wooldridge.
\newblock \emph{Econometric analysis of cross section and panel data}.
\newblock MIT press, 2010.

\bibitem[Yeager et~al.(2016)]{yeager2016using}
D.~S. Yeager et~al.
\newblock Using design thinking to improve psychological interventions: The
  case of the growth mindset during the transition to high school.
\newblock \emph{Journal of Educational Psychology}, 108\penalty0 (3):\penalty0
  374, 2016.

\bibitem[Zhao et~al.(2017)Zhao, Small, and Ertefaie]{zhao2017selective}
Q.~Zhao, D.~S. Small, and A.~Ertefaie.
\newblock Selective inference for effect modification via the lasso.
\newblock \emph{arXiv preprint arXiv:1705.08020}, 2017.

\end{thebibliography}

\end{document}